\documentclass[aps,prc,twocolumn,showpacs]{revtex4-1}
\usepackage{graphicx}
\usepackage{CJK}
\begin{document}
\begin{CJK*}{GBK}{song}
\title{On the Energy and Centrality Dependence of Higher Order Moments of  Net-Proton Distributions
in Relativistic Heavy Ion Collisions}
\author{ X. Wang   and C. B. Yang }
\affiliation{Institute of Particle Physics, Central China Normal
University,Wuhan 430079, People's Republic of China}
\affiliation{Key Laboratory of Quark and Lepton Physics (CCNU),
Ministry of Education, People's Republic of China}
\date{\today}
\begin{abstract}

  The higher order moments of the net-baryon distributions in relativistic
  heavy ion collisions are useful probes for the QCD critical point and fluctuations.
 Within a simple model we study the colliding energy and centrality dependence of the
 net-proton distributions in the central rapidity region. The model is based on considering
 the baryon stopping and pair
 production effects in the processes. Based on some physical reasoning, the dependence is
 parameterized. Predictions for the net-proton distributions for Au+Au and Pb+Pb collisions at different
 centralities at $\sqrt{s_{NN}}$=39 and 2760 GeV, respectively, are presented from the parameterizations for
 the model parameters. A possible test of our model is proposed from investigating the net-proton
 distributions in the non-central rapidity region for different colliding centralities and energies.

\end{abstract}
\maketitle

\section{Introduction}
The investigation of QCD phase diagram is of crucial importance for our understanding of the properties
of matter with strong interactions. Lattice QCD calculations have predicted,  at vanishing baryon
chemical potential, the occurrence of a cross-over from hadronic phase to the deconfined quark-gluon plasma
phase above a critical temperature of about 170-190 MeV \cite{YA,JB}. A distinct singular feature of the phase
diagram is the QCD critical point \cite{MAS} which is located at the end of the transition boundary.
A characteristic feature of the QCD critical point for systems in the thermodynamical limit is the divergence
of the correlation length $\xi$ and extremely large critical fluctuations. In ultra-relativistic heavy ion
 collisions, however, because of finite size and rapid expansion of the produced system, those divergence
 may be washed out. As estimated in \cite{MAS}, the critical correlation length in heavy ion collisions is
 not divergent but only about 2-3 fm. So the signals for the critical point of the system produced in
 heavy ion collisions cannot be observed as clearly as in the condensed matter physics.
However, remnants of those critical large fluctuations may become accessible in heavy ion collisions
through an event-by-event analysis of fluctuations in various channels of conservative  hadron quantum
numbers, for example, baryon number, electric charge, and strangeness \cite{FLUC}.
Particularly there would be a non-monotonic behavior of non-Gaussian multiplicity fluctuations
in an energy scan, which would be a clear signature for the existence of a critical point.
In fact, at vanishing chemical potential it has been shown that moments of conservative charge
distributions are sensitive indicators for the occurrence of a transition from hadronic to partonic matter \cite{SE}.

  Recently, great interest both experimentally \cite{STAR} and theoretically \cite{THEO1,THEO2,THEO3}
  has been aroused on the higher order moments of net-baryon distributions in heavy ion collisions at the BNL
  Relativistic Heavy Ion Collider (RHIC) energies.
  The theoretical interest on these higher order moments comes from the discovery of the relation
between the moments and the thermal fluctuations near the critical points for the produced quark matter.
If some memory of the large correlation length in the quark matter sate persists in
the thermal medium in hadronization process, this must be reflected in higher order moments of the
distributions. Theoretical prediction  \cite{MAS2} showed that the third moment, called skewness, is
proportional to $\xi^{4.5}$ and that the fourth moment, or kurtosis, proportional to $\xi^7$ while
  the second moment proportional to $\xi^2$. More importantly, the moments are closely related
  to the susceptibilities of the thermal medium. Thus the higher order moments have stronger dependence
  on the correlation length $\xi$ and are therefore more sensitive to the critical fluctuations. Recently
  STAR Collaboration has published experimental data on the higher order moments \cite{mom} for different
  colliding systems at different colliding energies for different colliding centralities. It has
  been argued that the net-proton distribution can be a meaningful observable for the
  purpose of detecting the critical fluctuations of net baryons in heavy ion collisions \cite{NP}.
  This statement makes experimental investigation of net-baryon fluctuations much easier, because
  neutrons and strange baryons cannot been detected easily/effectively in experiments.
 Based on theoretical and experimental investigations, it has been argued that information of QCD
  phase diagram  and the critical point  can be obtained from the energy dependence of those
  moments \cite{THEO1}.

  The moments of net-proton distributions have been studied recently by quite a few groups with different
 event generators such as A Multi-Phase Transport (AMPT) and Ultrarelativistic Quantum Molecular
 Dynamics (UrQMD) \cite{THEO3}, Heavy Ion Jet INteraction Generator (HIJING) \cite{LUO},
 and hadron resonance gas model \cite{HRG,FK} etc. Some other authors tried to search the statistical and
 dynamical components in the net-proton distributions in \cite{chen}, where the statistical distributions
 for both proton and anti-proton are assumed Poissonian and the departure from Poisson distributions
 is regarded as the dynamical influence. One should pay attention that the use of independent
 Poisson distributions for proton and anti-proton implies that protons and anti-protons are produced
 completely uncorrelated.
 Therefore the baryon number conservation may be violated in any event. On the multiplicity
 distribution of hadrons,  a canonical ensemble is employed in \cite{begun} to derive the number distribution
 for $\pi$ systems. This is reasonable because there are a lot of $\pi$ particles in the final state of
 heavy ion collisions. But a simple transportation of the method to the case for baryon production may
 be problematic, because the relevant baryon particle number may be not large enough for an
 equilibrium statistical description. In Ref. \cite{yw} the net-proton distributions in Au+Au collisions
 at $\sqrt{s_{NN}}=200 {\rm GeV}$ are studied from very simple but well established physics considerations:
 baryon stopping and  baryon pair production. For a given mean net-proton number from initial
 nuclear stopping, the initial proton number is assumed to satisfy a Poisson distribution.
 From the produced baryon pairs, the joint distribution for the newly produced proton and anti-proton
 can be derived. Then one can obtain the net-proton distribution in the final state of heavy ion collisions.
 Good agreement with experimental data has been obtained for Au+Au collisions at three colliding
 centralities at $\sqrt{s_{NN}}=200 {\rm GeV}$.

 This paper is an extension of the work in Ref. \cite{yw} by studying the moments of
 net-proton distributions in a given central rapidity window in Au+Au collisions at lower RHIC
 energies at different centralities.
  This paper is organized as follows. In next section, we will address our model and the physics points for
  the centrality dependence of parameters. Using an analytical expressions for the net-proton distribution
  derived in Ref. \cite{yw} we will show that the moments of the distribution up to fourth order
  can all be well described by our model with suitably chosen parameters for four colliding energies
  in Au+Au collisions at RHIC.  In section III, our model results for the moments are compared with the
  experimental data from STAR Collaboration. The net-proton distributions for Au+Au collisions at
  $\sqrt{s_{\rm NN}}$=39 GeV  are presented for three centralities. Then in section IV,  we discuss
  the energy dependence of the parameters and give parameterizations for such dependence.
  Then we extrapolate the dependence to the CERN Large Hadron Collider (LHC) energy and predict
  the net-proton distributions for Pb+Pb collisions at different colliding centralities for $\sqrt{s_{\rm NN}}$
  =2.76 TeV.  The last section will be for a brief summary.

\section{Model consideration for the centrality dependence}
Nuclear stopping plays an important role in heavy ion collisions and the study of such effect is a fundamental
issue, since this effect is related to the amount of energy and baryon number that get transferred
from the beam nucleons into the reaction zone.  We denote $B$ the mean net-proton number in the final state
 distribution. As can be seen from our model consideration, $B$ comes only from the initially stopped
 protons. In nuclear-nuclear collisions the mean net-baryon number $B$ in central rapidity region
would be zero if there were no nuclear stopping in the processes.
Another physics point we consider in Ref. \cite{yw} and here is baryon pair production in the interactions.
We denote $\mu$ the mean number of produced baryon pairs within a given kinematic region in the collisions
at given colliding centrality.
By assuming that baryon pairs are produced independently, the pair number distribution is of Poissonian. With
isospin conservation, one can derive a simple analytical formula for the net-proton number $\Delta p$
distribution $P(\Delta p)$ as a function of $B$ and $\mu$ as \cite{yw}
\begin{eqnarray}
P(\Delta p)&=&\int_0^\pi \frac{dx}{\pi} e^{-(2B+\mu)\sin^2\frac{x}{2}}\cos(x\Delta p-B\sin x)\ .\label{eq_sin}
\end{eqnarray}

Now we study the centrality dependence of $B$ and $\mu$.
Because the nuclear stopping effect results from the interactions between a passing nucleon with
other nucleons on its way in a nuclear-nuclear collision, the baryon number stopped in a given kinematic
region is closely related to the number of participant nucleons
$N_{\rm p}$. In more central collisions the stopped net baryon number
will be larger. In nuclear-nuclear collisions, if every nucleon from a nucleus suffers exactly the same
interactions, the stopped proton number would be $B\propto N_{\rm p}$. Of course, the real
case is not so simple. Because multiple scattering effect is more important for more central collisions,
a little larger $B/N_{\rm p}$ can be expected for more central collisions. On the other hand, the stopped
net-proton can be detected only in the final state, or in other words, after evolving with the system
for some time. During the evolution of the system,
the net-proton may diffuse into a kinematic region out of our interested window. For central collisions,
the evolution time is longer and such diffusion effect is more obvious. This effect would reduce  $B/N_{\rm p}$
 for central collisions. The real proportional factor  $B/N_{\rm p}$ is a result from
the competition of the two effects: initial multiple scattering and later baryon number diffusing. Therefore,
the factor $B/N_{\rm p}$ should have a weak centrality dependence. Thus one may parameterize the centrality
dependence of $B$ as
\begin{equation}
B=a_1N_p(1-a_2N_p)\ ,
\end{equation}
with $a_1$ and $a_2$ depending on the colliding energy of the system. While $a_1$ is always positive,
the magnitude of $a_2$ should be small and $a_2$ can be negative or positive, depending on whether
multiple scattering is more important than
baryon number diffusion or the opposite. For the baryon pair production, similar physics considerations
apply also. The probability for a baryon pair production near a point in the system is determined by
the energy density at that point. If the energy density of the hot strong interacting matter is
uniform and the same for all colliding
centralities, one may expect $\mu$ proportional to the volume of the system, thus $\mu\propto N_{\rm p}$.
But the initial energy density is higher for more central collisions due to stronger nuclear stopping effect,
a larger $\mu/N_{\rm p}$ is expected for more central collisions,
if the nuclear stopping is the only physics in the process. On the other hand,
as in the consideration for $B$, the energy diffusion from central to non-central
 rapidity region will reduce the value of $\mu/N_{\rm p}$.
The competition of these two effect results in a behavior of $\mu$ as a function of $N_{\rm p}$ similar
to that of $B$. Therefore we parameterize the colliding centrality dependence of $\mu$ as
\begin{equation}
\mu=b_1N_p(1-b_2N_p)\ ,
\end{equation}
with $b_1$ and $b_2$ also depending on the colliding energy. As for $a_2$, the value of $b_2$ should be
very small but can be positive or negative due to the competition of initial energy stopping and
diffusion in the evolution of the system.

With the above four parameters, one can calculate the net-proton distribution and all the associate
moments for any colliding centrality for given center of mass energy of the colliding system.

\section{Comparison with the experimental Data}
For Au+Au collisions at RHIC energies, we investigate the moments up to fourth order for the distribution
of net-proton in the central rapidity window $|y|<0.5$ as functions of
$N_{\rm p}$ by using our model described in the last section. The fitted results from our model for
the moments are shown in Figs. 1-4 for the mean, variance, skewness and kurtosis for four colliding
energies $\sqrt{s_{\rm NN}}$=19.6, 39, 62.4 and 200 GeV. The fitted parameters are tabulated in
TABLE I. It can be seen that our simple model can describe quite well the centrality dependence
of moments for the four energies with the parameters chosen. From the excellent agreement with the
experimental data, one can conclude that our model contains the necessary physics for the net-proton
distributions.

\begin{figure}[tbph]
\includegraphics[width=0.45\textwidth]{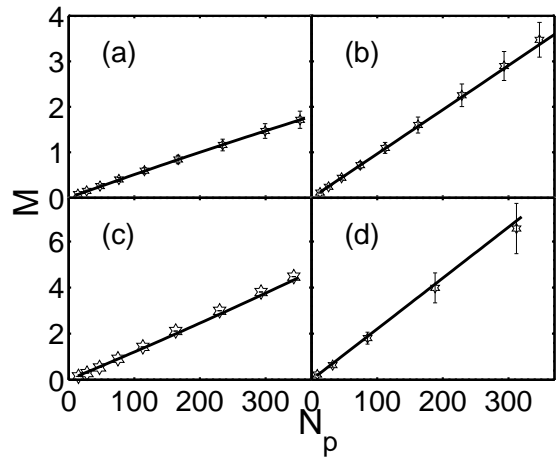}
\caption{Mean value of the net-proton distributions as a function of $N_{\rm p}$ for Au+Au collisions
at four $\sqrt{s_{\rm NN}}$, (a) 200; (b) 62.4; (c) 39 and (d) 19.6 GeV. The points are from RHIC/STAR data
\cite{mom}. }
\end{figure}

\begin{figure}[tbph]
\includegraphics[width=0.45\textwidth]{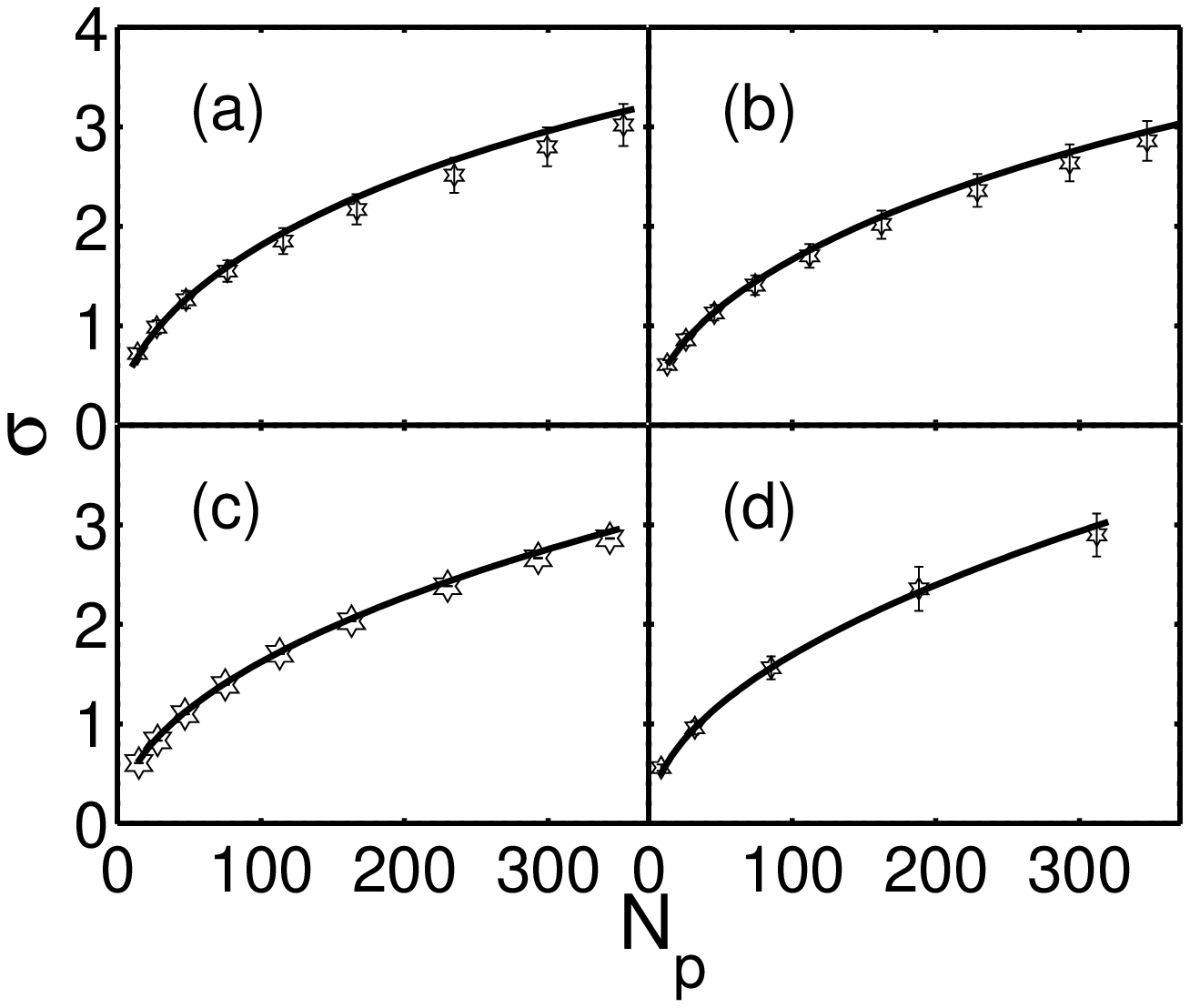}
\caption{variance of the net-proton distributions as a function of $N_{\rm p}$ for Au+Au collisions
at four $\sqrt{s_{\rm NN}}$, (a) 200; (b) 62.4; (c) 39 and (d) 19.6 GeV. The points are from RHIC/STAR
data \cite{mom}.}
\end{figure}

\begin{figure}[tbph]
\includegraphics[width=0.45\textwidth]{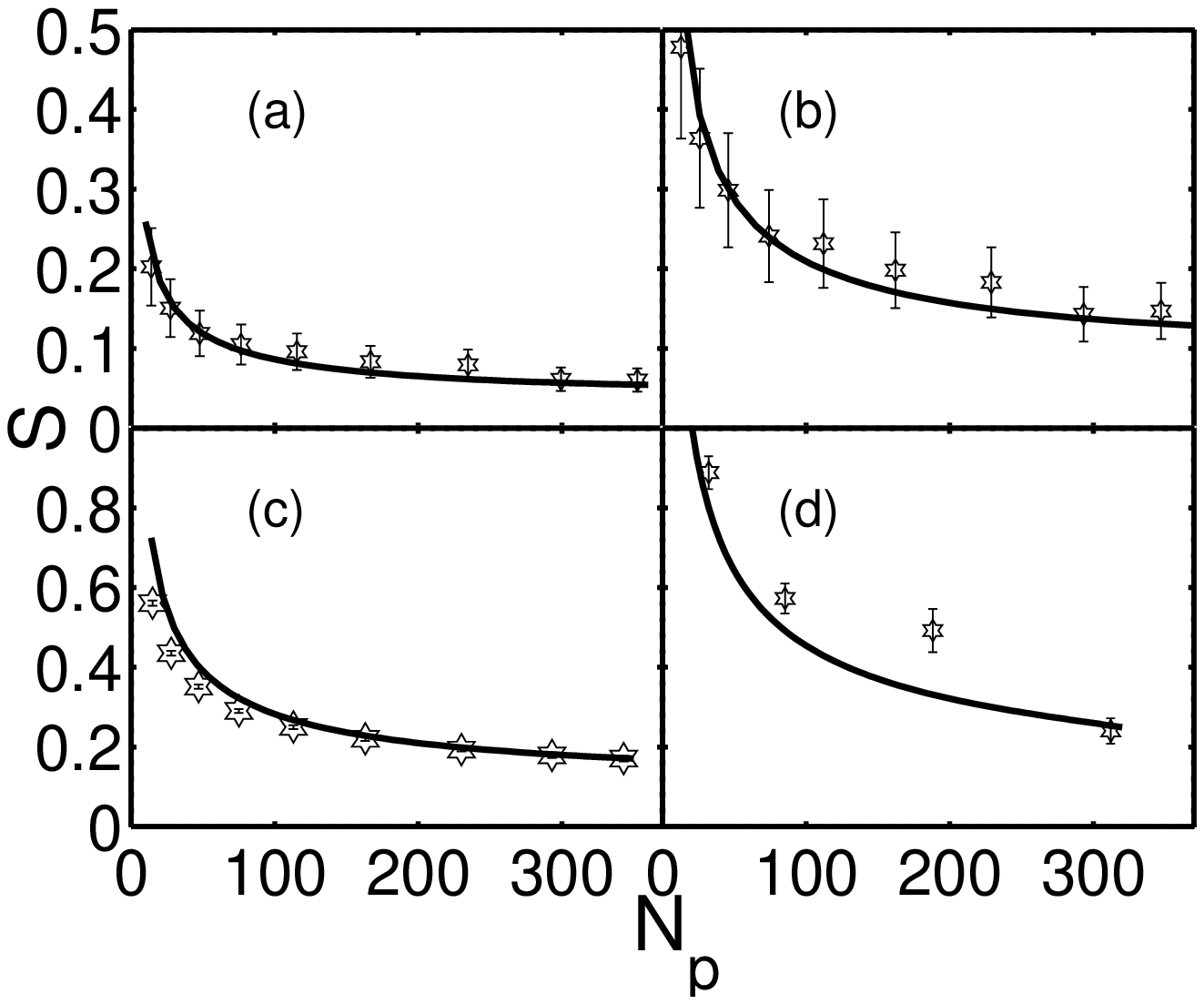}
\caption{Skewness of the net-proton distributions as a function of $N_{\rm p}$ for Au+Au collisions
at four $\sqrt{s_{\rm NN}}$, (a) 200; (b) 62.4; (c) 39 and (d) 19.6 GeV.
The points are from RHIC/STAR data \cite{mom}.}
\end{figure}

\begin{figure}[tbph]
\includegraphics[width=0.45\textwidth]{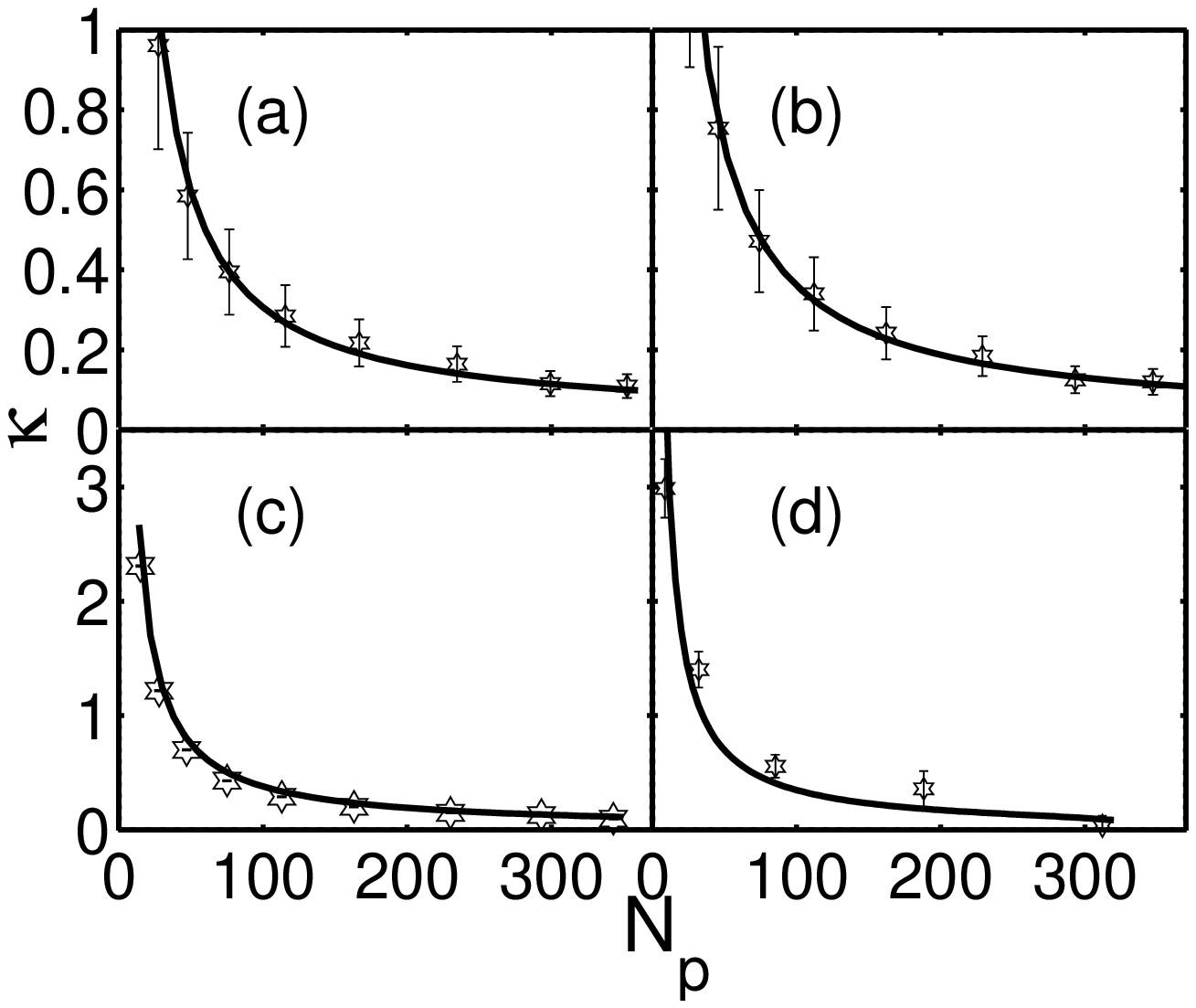}
\caption{Kurtosis of the net-proton distributions as a function of $N_{\rm p}$ for Au+Au collisions
at four $\sqrt{s_{\rm NN}}$, (a) 200; (b) 62.4; (c) 39 and (d) 19.6 GeV.
The points are from RHIC/STAR data \cite{mom}.}
\end{figure}

\begin{table}
\begin{tabular}{||c|c|c|c|c||}
\hline\hline
$\sqrt{s_{\rm NN}}$ (GeV) & $a_1$ & $10^4a_2$ & $b_1$ & $10^4b_2$\\ \hline
19.6 & 0.022 & $-0.126$ & 0.0132 & $0.121$\\ \hline
39 & 0.0012 & $-1.97$ & 0.0300 & $4.90$\\ \hline
62.4 & 0.0096 & $-0.159$ & 0.0383 & $5.10$\\ \hline
200 & 0.0052 & $1.88$ & 0.0585 & $5.00$ \\ \hline\hline
\end{tabular}
\caption{Fitted parameters for four colliding energies.}
\label{tab1}
\end{table}

With those parameters in TABLE \ref{tab1}, one can calculate the net-proton distributions
quite easily for different centralities for the four energies as in TABLE \ref{tab1}.
Our new parametrization for $B$ and $\mu$ can give distributions for Au+Au collisions at
$\sqrt{s_{\rm NN}}$=200 GeV. The newly obtained distributions have no visible difference from those in
Ref. \cite{yw}, in good agreement with the STAR data, so will not be presented here. As an example to
show the distributions, we present here only the distributions at $\sqrt{s_{\rm NN}}=39$ GeV for
Au+Au collisions at different colliding centralities, for later comparison with experimental results.
\begin{figure}[tbph]
\includegraphics[width=0.45\textwidth]{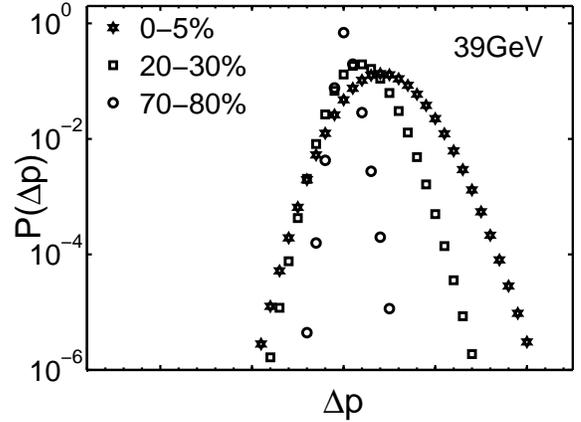}
\caption{Net-proton distributions for Au+Au collisions at $\sqrt{s_{\rm NN}}=39$ GeV at
different colliding centralities.}
\end{figure}

\section{Colliding energy dependence of net-proton distribution}
After discussing the net-proton distributions for Au+Au collisions at four different colliding
energies, one can discuss the colliding energy dependence of parameters in our model.
As we discussed in Sec. II, the values and their dependence on the colliding energy can tell us
some physics in the colliding process, especially the competition of effects from the initial multiple
scattering and later baryon number transportation. With the increase of colliding energy, the stopped
baryon number in the central rapidity region will decrease. Thus the
value of $a_1$  will decrease with $\sqrt{s_{\rm NN}}$. Though the energy fraction stopped
in the central rapidity region becomes smaller at higher colliding energy, the energy density
in that region still increases with $\sqrt{s_{\rm NN}}$. Then $b_1$ will increase with $\sqrt{s_{\rm NN}}$.
For $\sqrt{s_{\rm NN}}$ high enough, one may expect $a_1$ and $b_1$ saturates at some limiting values.
The behaviors of $a_2$ and $b_2$
can be quite different from those of $a_1$ and $b_1$. Since $a_2$ and $b_2$ depend on the competition
of effects from initial multiple nucleon-nucleon scattering and later baryon/energy diffusion, the behaviors
of $a_2$ and $b_2$ with the increase of $\sqrt{s_{\rm NN}}$ can be complicated. Both the effects from
the initial multiple nucleon-nucleon scattering and later baryon/energy diffusion become stronger with
the increase of $\sqrt{s_{\rm NN}}$, but their rates of increase may be different. For a given increase of
$\sqrt{s_{\rm NN}}$, if the initial multiple scattering becomes more important, $|a_2|$ ($|b_2|$) will
increase with $\sqrt{s_{\rm NN}}$ in the region with  negative $a_2$ ($b_2$). Otherwise,  $|a_2|$ ($|b_2|$)
will decrease in the same region.
 For $\sqrt{s_{\rm NN}}\to\infty$, the interaction duration in the produced matter
will very long and the diffusion effect will be much stronger than that from initial multiple scattering.
Then one can expect $a_2$ and $b_2$ approaching some positive saturating values, implying that $B/N_{\rm p}$
and $\mu/N_{\rm p}$ is smaller for central collisions when $\sqrt{s_{\rm NN}}$ is high enough.

\begin{figure}[tbph]
\includegraphics[width=0.45\textwidth]{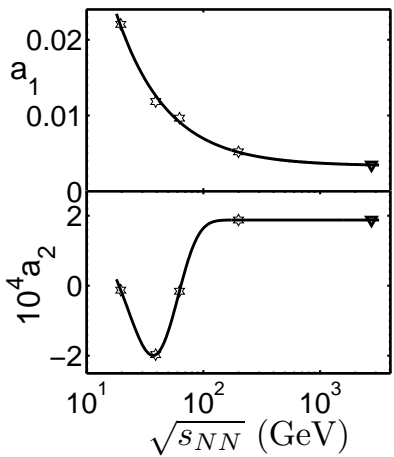}
\caption{Colliding energy dependence of parameters $a_1$ and $a_2$ for the initially stopped
proton number in the given central rapidity window. Points marked by star are from our model fitting,
and the points marked by triangle are calculated from Eq. (4) at $\sqrt{s_{\rm NN}}$=2760 GeV.}
\label{br}
\end{figure}

\begin{figure}[tbph]
\includegraphics[width=0.45\textwidth]{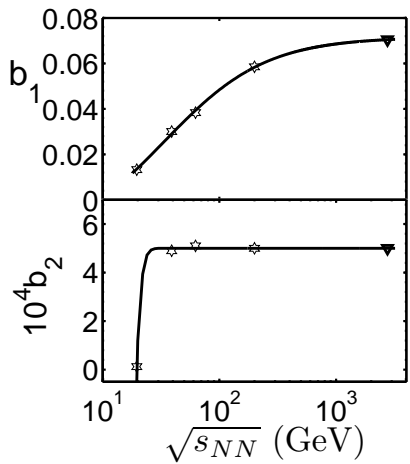}
\caption{Colliding energy dependence of parameters $b_1$ and $b_2$ for the mean number of produced baryon
 pairs in the given central rapidity window. Points marked by star are from our model fitting, and the points
marked by triangle are calculated from Eq. (4) at $\sqrt{s_{\rm NN}}$=2760 GeV. }
 \label{mu}
\end{figure}

\begin{figure}[tbph]
\includegraphics[width=0.45\textwidth]{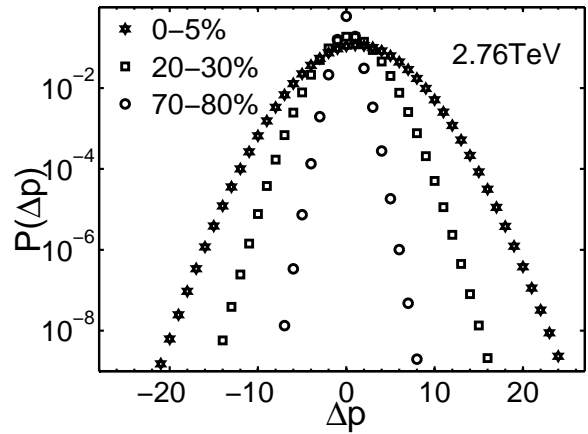}
\caption{Net-proton distributions for Pb+Pb collisions at $\sqrt{s_{\rm NN}}=2760$ GeV at
different colliding centralities. }
\label{lhc_dis}
\end{figure}

To see the colliding energy dependence of the four parameters, we plot the fitted parameters
shown in TABLE I as functions of $\sqrt{s_{\rm NN}}$ in GeV. The plots are shown in Figs.
 \ref{br} and \ref{mu}. The shown dependence of the parameters on the colliding energy
 can be described by the following expressions
\begin{equation}
\begin{array}{ccl}
a_1&=&14.55(1+9.23\times 10^{-3}\sqrt{s_{\rm NN}})/(1+40.2\sqrt{s_{\rm NN}})\ ,\\
10^4a_2&=&1.87-1.21\times 10^{-4}(\sqrt{s_{\rm NN}})^4\exp(-0.107\sqrt{s_{\rm NN}})\ ,\\
b_1&=&-0.02(1-0.105\sqrt{s_{\rm NN}})/(1+0.0295\sqrt{s_{\rm NN}})\ ,\\
10^4b_2&=& 5.0-0.626(\sqrt{s_{\rm NN}})^{7.14}\exp(-0.979\sqrt{s_{\rm NN}})\ . \\
\end{array}
\end{equation}
\noindent
The functional form for $a_i$ and $b_i$ ($i=1,2$) are chosen to satisfy the demands from
 the physics considerations in the last paragraph. With the above expressions for the energy dependence
 of the parameters, it is straightforward
to calculate the values of those parameter at the LHC energy $\sqrt{s_{\rm NN}}$=2760 GeV. By assuming
that our model can be applied to that energy, one can calculate the net-proton distributions
at that energy for different colliding centralities. The obtained distributions are shown in
Fig. \ref{lhc_dis}. The values of $N_{\rm p}$ used in the calculation are from \cite{LHC}.

In all considerations up to now, we only discussed the net-proton distributions in the central rapidity
region $|y|<0.5$. Because of the diffusion of baryon number and energy from central to non-central
rapidity region, the parameters $a_2$ and $b_2$ show complicated behaviors as functions of colliding
energy. If we consider the net-proton distributions in the non-central region, $|y|>0.5$ for say,
the same physics arguments apply and one can expect that the distributions can also be described by
Eq. (1) with centrality dependence of parameters $B$ and $\mu$ being given by Eqs. (2) and (3). One can
also expect that the colliding energy dependence of $a_1$ and $b_1$ is similar to that for the case
in central rapidity region. For $a_2$ and $b_2$, the baryon number and energy diffusion from central to
non-central region will not compete to but cooperate with the multiple scattering effect. Then $a_2$ and $b_2$ will be negative for all colliding energies and all A+A collisions.
 This statement can be tested experimentally.

\section{Conclusion}

We studied the net-proton distributions for Au+Au collisions at four colliding energies for
$\sqrt{s_{\rm NN}}$ from 19.6 to 200 GeV at different centralities. Based on some physical arguments,
the parameters in our model are parameterized as functions of centrality and energy. The higher order
moments for the distributions are in good agreement with the experimental data. Prediction for the net-proton
distributions at LHC energy $\sqrt{s_{\rm NN}}$=2.76 TeV is presented for different centralities. The net-proton
distribution in a non-central rapidity region and its dependence on centrality are discussed.

It should be mentioned that nothing else is assumed in this model except an initial stopped net-proton and a
finite probability for producing baryon pairs from the produced matter. Therefore, our model has nothing to do
with  thermal equilibrium and/or critical fluctuations. Because our model consideration is
based on normal physics effects, our results can be used as a baseline for detecting novel physics in the
processes.

\acknowledgments{This work was supported in part by the National Natural Science Foundation of
China under Grant No. 11075061 and by the Programme of Introducing Talents of Discipline to
Universities under No. B08033. The authors thank Dr. X.F. Luo for sending us the experimental data.
We are grateful to N. Xu and X.F. Luo for valuable discussions. }

\end{CJK*}
\end{document}